\documentclass[twocolumn,prl,showpacs,amsmath,amssymb]{revtex4}

\usepackage{graphicx}
\usepackage{dcolumn}
\usepackage{bm}

\begin{document}

\title{Experimental study of isovector spin sum rules}

\author{
A.~Deur$^{\njlab}$,
P.~Bosted$^{\njlab }$,
V.~Burkert$^{\njlab}$,
D.~Crabb$^{\nuva}$,
V.~Dharmawardane$^{\nodu}$\footnote{Present address: 
New Mexico State University,
Las Cruces, NM 88003},
G.~E.~Dodge$^{\nodu}$, 
T.~A.~Forest$^{\istu}$, 
K.~A.~Griffioen$^{\nwm}$, 
S.~E.~Kuhn$^{\nodu}$, 
R.~Minehart$^{\nuva}$, 
Y.~Prok$^{\nuva}$\footnote{Present address: Christopher Newport University, 
Newport News, VA 23606}.}

\affiliation{
\baselineskip 2 pt
\centerline{{$^{\njlab}$Thomas Jefferson National Accelerator Facility, 
Newport News, VA 23606}}
\centerline{{$^{\nuva}$University of Virginia, Charlottesville, VA 22904}}
\centerline{{$^{\nodu}$Old Dominion University,  Norfolk, VA 23529}}
\centerline{{$^{\istu}$Idaho State University, Pocatello, ID, 83209}}
\centerline{{$^{\nwm}$College of William and Mary, Williamsburg, VA 23187}}
}

\newcommand{\nltec}{4}
\newcommand{\nodu}{3}
\newcommand{\njlab}{1}
\newcommand{\nuva}{2}
\newcommand{\istu}{4}
\newcommand{\nwm}{5}

\date{\today}

\begin{abstract}
We present the Bjorken integral extracted from Jefferson Lab
experiment EG1b for $0.05<Q^{2}<2.92$ GeV$^2$. The integral is fit 
to extract the twist-4 element $f_{2}^{p-n}$ which appears to be 
relatively large and negative. Systematic studies of this higher 
twist analysis establish its legitimacy at $Q^{2}$ around 1 
GeV$^{2}$. We also performed an isospin decomposition of 
the generalized forward spin polarizability $\gamma_{0}$. Although 
its isovector part provides a reliable test of the calculation
techniques of Chiral Perturbation 
Theory, our data disagree with the calculations. 
\end{abstract}


\maketitle

\section{Introduction}

The Bjorken sum rule~\cite{Bjorken SR} relates an integral over the 
spin distributions of quarks inside the nucleon to its axial charge. 
This relation has been essential for understanding the 
nucleon spin structure and establishing, \emph{via} its 
$Q^2$-dependence, that Quantum Chromodynamics (QCD) describes 
the strong force when spin is included.
The Bjorken integral has been measured in polarized deep inelastic
lepton scattering (DIS) at SLAC, CERN and DESY~\cite{SLAC}-\cite{HERMES}
and at moderate four-momentum transfer squared $Q^2$
at Jefferson Lab (JLab)~\cite{EG1a/E94010}, see e.g. Ref.~\cite{review} for a review. The
variable $Q^2$ is inversely related to the space-time scale at which 
the nucleon is probed. In the perturbative QCD (pQCD) domain (high 
$Q^2$) the sum rule reads~\cite{Bj RadCor}:
\begin{eqnarray}
\label{eq:bj(Q2)}
\Gamma_{1}^{p-n}(Q^2)\equiv\int_{0}^{1}dx
\left( g_{1}^{p}(x,Q^2)-g_{1}^{n}(x,Q^2) \right)=
\hspace{1cm}\\
\frac{g_{A}}{6}\left[1-\frac{\alpha_{s}}{\pi}-3.58
\frac{\alpha_{s}^{2}}{\pi^{2}}-
20.21\frac{\alpha_{s}^{3}}{\pi^{3}}+...\right]+
\sum_{i=2}^{\infty}{\frac{\mu_{2i}^{p-n}(Q^{2})}{Q^{2i-2}}} \nonumber
\end{eqnarray}
where $g_{1}^{p}$ and $g_{1}^{n}$ are the spin-dependent proton
and neutron structure functions, $g_{A}$ is the nucleon axial charge 
that controls the strength of neutron $\beta$-decay, $\alpha_s(Q^2)$ 
is the strong coupling strength and $x=Q^{2}/2M\nu$, with $\nu$ the 
energy transfer and $M$ the nucleon mass. The bracket term ($\mu_2$, 
known as the leading twist term) is mildly dependent on $Q^2$ due
to pQCD soft gluon radiation. The other term contains non-perturbative 
power corrections (higher twists). These are quark and gluon 
correlations that need to be understood to describe the nucleon structure
away from the large $Q^2$ limit. The $Q^{2}$-dependence of 
$\mu_{2i}(Q^{2})$ is calculable in principle from pQCD. In practice, 
this has been done for $\mu_2$ and $\mu_4$ only~\cite{shuryak}. 
We stress that, as is almost always the case with pQCD, although the 
$Q^{2}$-dependences are known, the absolute values of $\mu_2$ and $\mu_4$
are unknown and need to be measured or computed by non-perturbative 
means. Besides its contribution to establishing pQCD (at high $Q^2$), 
the Bjorken sum rule can be used to extract higher twists, to check 
lattice QCD calculations (at moderate $Q^2$), and to test effective 
theories of the strong force (at low $Q^2$).  In addition, Bjorken 
sum data and phenomenological models at lower $Q^2$ can be described 
with a nearly constant ``effective strong coupling'' 
$\alpha_{s,g1}$~\cite{alpha_s_eff,alpha_s_eff 2}.
The lack of $Q^2$-dependence of  $\alpha_{s,g1}$ opens new avenues for 
non-perturbative QCD calculations using the AdS/CFT 
correspondence~\cite{ads/CFT}.

The elastic contribution to the Bjorken sum is usually not included
because the generalized Bjorken sum rule is derived at large $Q^2$ where 
such contribution is negligible. Furthermore, the Bjorken sum rule 
naturally connects to the Gerasimov-Drell-Hearn sum rule~\cite{GDH} in 
which the elastic is inexistent.
Consequently, when presenting the experimental measurement of the 
Bjorken sum, the elastic contribution will not be included. We refer to
ref.~\cite{X Ji. elastic} for a discussion on whether to include 
or not the elastic contribution to the GDH sum rule. However, for 
higher twist analysis, all reactions should be included for a
meaningful higher twist extraction~\cite{X Ji. elastic}-\cite{Musatov}. 
Therefore, in the part of the paper discussing higher twist
extraction, the elastic contribution to the Bjorken sum will be added.

In this paper, new data from the JLab CLAS EG1b 
experiment~\cite{EG1b}-\cite{EG1b moments} 
taken on polarized proton and deuteron targets are used 
to extract the Bjorken integral over an extended $Q^{2}$ range:
$0.05<Q^2<2.92$ GeV$^2$ compared to the previous JLab range
$0.15<Q^2<1.5$ GeV$^2$~\cite{EG1a/E94010}.

The extension down to $Q^{2}=0.05$ GeV$^2$ allows us to compare to 
Chiral Perturbation Theory ($\chi PT$) calculations in a domain where 
the chiral 
approximation should be valid. The moderate $Q^{2}$ range had been 
precisely measured~\cite{EG1a/E94010}. The new data set, of equivalent 
precision, provides a useful check. In particular, it verifies the 
neutron results, which come mostly from $^{3}$He in 
Ref.~\cite{EG1a/E94010} and from the deuteron in this paper. 
At larger $Q^{2}$ ($\gtrsim$ 1 GeV$^2$, where Eq.~\ref{eq:bj(Q2)} 
holds), higher twists can now be studied with a statistical precision 
typically improved by a factor of 2. Previous work~\cite{EG1a/E94010}
has shown the necessity of precise $Q^{2}$ mapping at moderate
$Q^{2}$ ($\gtrsim$ 1 GeV$^2$) because of the surprisingly small size 
of the overall higher twist effect. One might be tempted to lower the 
$Q^{2}$ values at which the analysis is done (see Eq.~\ref{eq:bj(Q2)}) 
but this is not reliable due to the fast $1/Q^{2i-2}$ rise of 
twist $i$ contributions 
at low $Q^{2}$ and to the increasing uncertainty of the evolution of the 
twist-2 parts. The main contributor at low $Q^2$ to this uncertainty 
is the strong coupling constant $\alpha_s(Q^2)$.

The Bjorken integral is advantageous compared to
the individual moments $\Gamma_1^p$ and $\Gamma_1^n$ because of 
simplifications arising from its non-singlet (p-n) nature: at
moderate $Q^2$ lattice QCD calculations are easier and more reliable 
because disconnected diagrams, which cannot be easily computed on the 
lattice, cancel out. At higher $Q^2$, the (p-n) simplification provides 
a sum rule (the Bjorken sum rule)  based on more solid grounds  than the 
sum rules for individual nucleons (the Ellis Jaffe sum rules~\cite{EJ} 
that necessitate additional assumptions). At low $Q^2$, the (p-n) 
subtraction cancels the $\Delta_{1232}$ resonance contribution which 
makes the $\chi PT$ calculations significantly more 
reliable~\cite{Burkert Delta}. By a similar argument, the 
transverse-longitudinal polarizability $\delta_{LT}$~\cite{review}, 
a higher moment of spin structure functions, also provides a reliable 
test of $\chi PT$ computations. (In that case, the $\Delta_{1232}$ 
contribution
is suppressed at low $Q^2$ because the N-$\Delta$ transition is mostly
transverse, making the longitudinal-transverse (LT) interference term 
very small.) Nevertheless, calculations based on $\chi PT$  and data 
for $\delta_{LT}$ 
on  the neutron~\cite{E94010 delta_lt} strongly disagree. This 
calls for more low $Q^2$ studies, especially the yet unmeasured 
$\delta_{LT}^p$~\cite{delta_lt prop}. The data discussed in this paper
were taken with a 
longitudinally polarized target and hence cannot be used to extract 
$\delta_{LT}^{p}$. However, the generalized forward spin 
polarizability $\gamma_{0}$ can be obtained and, just like the 
Bjorken integral, its isovector part $\gamma_{0}^{p}-\gamma_{0}^{n}$ 
offers the same advantages as $\delta_{LT}$ for checking the 
calculation techniques of $\chi PT$. 
We will also report on these results.

\section{Bjorken sum extraction and comparison with Chiral Perturbation
calculations }
The measurements of structure functions $g_{1}^{p}$ and $g_{1}^{d}$
are described in Refs.~\cite{EG1b}-\cite{EG1b moments}. The data cover 
an invariant mass range up to $W=$ 3 GeV for 0.054 $\leq Q^{2} \leq$ 
2.92 GeV$^{2}$. Since experimental moments are integrated over a 
finite $W$ range, the data have to be supplemented by models for
large $W$. We used the model described in Ref.~\cite{EG1b}
down to $x=0.001$. This part is known from DIS experiments. 
The rest is determined using a Regge parametrization~\cite{EG1a/E94010}  
which was compared to that of Bass and Brisudova~\cite{Bass low-x} and 
found consistent with it. A parameterization was also used to estimate 
the contributions between pion threshold (1.08 GeV) and 1.15 
GeV~\cite{EG1b}.

\begin{figure}[ht!]
\begin{center}
\vspace*{-1.0cm}
\centerline{\includegraphics[scale=0.48, angle=0]{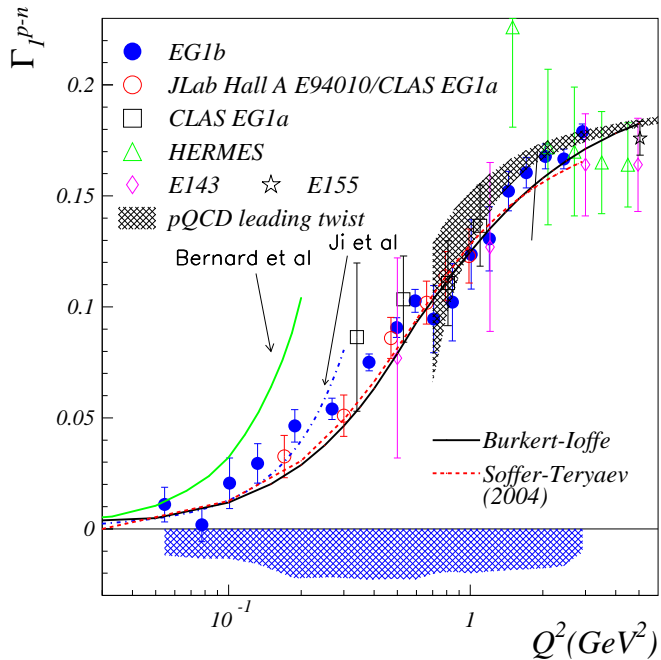}}
\end{center}
\vspace{-1.0cm}
\caption{(Color online) The Bjorken integral $\Gamma_1^{p-n}(Q^2)$. 
The solid blue circles give the results from this work with the 
horizontal band giving the systematic uncertainties. Other symbols 
show the data from experiments 
E143~\cite{E143} (open diamonds), 
E155~\cite{E155-E155x} (open star),
HERMES~\cite{HERMES} (open triangles) 
and JLab~\cite{EG1a/E94010} (open circles and open squares). 
For those, the error bars represent the quadratic sum of the statistic 
and systematic uncertainties. The gray band represents the 
leading-twist NNLO pQCD calculation. The curves correspond to 
$\chi PT$ calculations (~\cite{meissner, Ji chipt}) and phenomenological 
models (~\cite{AO, soffer}).}
\label{fig:bjsr}
\end{figure}

The Bjorken integral is obtained from $\Gamma_{1}^{p}$ and
$\Gamma_{1}^{d}$ assuming:
\vspace{-0.2cm}
\begin{eqnarray*}
\Gamma_{1}^{p-n}=2\Gamma_{1}^{p}-\Gamma_{1}^{d}/\left(1-1.5\omega_{d}\right),
\end{eqnarray*}
with the deuteron D-state probability 
$\omega_{d}=0.05\pm0.01$~\cite{omega_d}. The data are given in Table I 
(a more detailed table is given in~\cite{EG1b bjorken note}) and shown in 
Fig.~\ref{fig:bjsr}. The elastic contribution 
($x = 1$) is excluded. Data from SLAC E143~\cite{E143}, 
HERMES~\cite{HERMES}, JLab CLAS EG1a (proton and deuteron), and JLab 
Hall A E94010 (neutron from $^{3}$He) combined with CLAS EG1a 
(proton)~\cite{EG1a/E94010} are also shown for comparison.

\begin{table}
\begin{tabular}{|c|c|c|c|c|c|}
\hline 
{\large $Q^{2}$ }&
{\large $\Gamma_{1,meas}^{p-n}$}&
{\large $\sigma_{meas}^{syst}$}&
{\large $\Gamma_{1,tot}^{p-n}$}&
{\large $\sigma^{syst}$}&
{\large $\sigma^{stat}$}\tabularnewline
\hline
\hline 
{ 0.054}&
{ 0.0028}&
{ 0.0105}&
{ 0.0110}&
{ 0.0119}&
{ 0.0078}\tabularnewline
\hline 
{ 0.078 }&
{ -0.0085}&
{ 0.0112}&
{ 0.0019}&
{ 0.0134}&
{ 0.0076}\tabularnewline
\hline 
{ 0.101 }&
{ 0.0076 }&
{ 0.0105}&
{ 0.0206}&
{ 0.0134}&
{ 0.0114}\tabularnewline
\hline 
{ 0.132 }&
{ 0.0129}&
{ 0.0124}&
{ 0.0296}&
{ 0.0158}&
{ 0.0089}\tabularnewline
\hline 
{ 0.188 }&
{ 0.0209}&
{ 0.0181}&
{ 0.0464}&
{ 0.0223}&
{ 0.0073}\tabularnewline
\hline 
{ 0.268 }&
{ 0.0155}&
{ 0.0152}&
{ 0.0541}&
{ 0.0218}&
{ 0.0048}\tabularnewline
\hline 
{ 0.382 }&
{ 0.0197}&
{ 0.0139}&
{ 0.0750}&
{ 0.0229}&
{ 0.0038}\tabularnewline
\hline 
{ 0.496 }&
{ 0.0184}&
{ 0.0110}&
{ 0.0907}&
{ 0.0225}&
{ 0.0045}\tabularnewline
\hline 
{ 0.592 }&
{ 0.0318}&
{ 0.0143}&
{ 0.1027}&
{ 0.0228}&
{ 0.0052}\tabularnewline
\hline 
{ 0.707 }&
{ 0.0513}&
{ 0.0174}&
{ 0.0945}&
{ 0.0201}&
{ 0.0151}\tabularnewline
\hline 
{ 0.844 }&
{ 0.0507}&
{ 0.0157}&
{ 0.1021}&
{ 0.0193}&
{ 0.0174}\tabularnewline
\hline 
{ 1.01 }&
{ 0.0656}&
{ 0.0152}&
{ 0.1236}&
{ 0.0200}&
{ 0.0156}\tabularnewline
\hline 
{ 1.20}&
{ 0.0628}&
{ 0.0161}&
{ 0.1307}&
{ 0.0192}&
{ 0.0145}\tabularnewline
\hline 
{ 1.44 }&
{ 0.0718}&
{ 0.0141}&
{ 0.1522}&
{ 0.0186}&
{ 0.0089}\tabularnewline
\hline 
{ 1.71 }&
{ 0.0695}&
{ 0.0129}&
{ 0.1605}&
{ 0.0182}&
{ 0.0069}\tabularnewline
\hline 
{ 2.05 }&
{ 0.0616}&
{ 0.0118}&
{ 0.1678}&
{ 0.0177}&
{ 0.0056}\tabularnewline
\hline 
{ 2.44 }&
{ 0.0458}&
{ 0.0098}&
{ 0.1666}&
{ 0.0167}&
{ 0.0045}\tabularnewline
\hline 
{ 2.92}&
{ 0.0483}&
{ 0.0079}&
{ 0.1789}&
{ 0.0106}&
{ 0.0035}\tabularnewline
\hline
\end{tabular}

\caption{The measured ($\Gamma_{1,meas}^{p-n}$) and total 
($\Gamma_{1,tot}^{p-n}$) Bjorken integrals for different $Q^2$ points 
(in GeV$^2$). The experimental systematic uncertainty 
$\sigma_{meas}^{syst}$ is given in the 3$^{rd}$ column. Total 
systematics uncertainty, including the low and large-$x$ extrapolations, 
($\sigma^{syst}$) and statistical uncertainty ($\sigma^{stat}$) on 
$\Gamma_{1,tot}^{p-n}$ are given in the $5^{th}$ and $6^{th}$ columns.}

\end{table}

There is excellent agreement between the Bjorken integral with the 
neutron extracted from the deuteron (filled circles and open
squares) and  from $^{3}$He (open circles). The neutron spin structure 
functions extracted from the deuteron and from
$^3$He agree at moderate and large $Q^2$. However, for $Q^2$ 
below a few tenths of a GeV$^2$, nuclear effects beyond those accounted 
for in the convolution method employed to extract the
neutron~\cite{degli_atti} may become large~\cite{low Q neutron}.
Therefore, at low $Q^2$ one needs both the deuteron and $^3$He data 
to ensure a reliable neutron extraction. Nuclear effects in the 
deuteron are weaker, but there 
is an unsuppressed contribution from the proton. On the other hand, 
$^3$He is more tightly bound, but the polarized proton contribution is 
largely suppressed. Consequently, the uncertainty due to nuclear effects 
is mostly of different origin in the deuteron and $^3$He, which 
makes the two nuclei complementary. The agreement 
between the deuteron and $^3$He results is also encouraging for the 
interpretation of the low $Q^2$ $^3$He and the deuteron data 
($Q^2>0.015$ GeV$^2$) that will be available shortly respectively from 
Jefferson Lab's Hall A~\cite{E97110} and B~\cite{EG4D}. 
The data also agree well with the SLAC and HERMES
experiments and with the two phenomenological models shown in
Fig.~\ref{fig:bjsr}. The model of Burkert and Ioffe~\cite{AO} 
(continuous black curve) is a meson-dominance-based extrapolation of 
DIS data supplemented by a parametrization of the resonance 
contribution. The other model (Soffer-Teryaev~\cite{soffer}, dashed red 
curve) uses the smoothness of $g_{1}+g_{2}$ with $Q^{2}$ to extrapolate 
DIS data at lower $Q^{2}$.

At moderate $Q^{2}$, we observe a strong variation of the Bjorken
integral, in contrast to the high $Q^{2}$ region. Together with our data
at the lowest $Q^2$ points, the kinematic constraint 
$\Gamma_1 \rightarrow 0$ when $Q^2 \rightarrow 0$ suggests a small 
$Q^2$-dependence of $\Gamma_1^{p-n}$ at low $Q^2$ as well.
This would agree with the fact that the $\Gamma_1$ slope at 
$Q^2 \simeq 0$ is given by the generalized GDH sum rule which predict 
a small $Q^2$-dependence.

At low $Q^{2}$ the data are consistent up to $Q^{2}$$\simeq$ 0.2 
GeV$^{2}$ with the $\chi PT$ calculations of Bernard 
\emph{et al.}~\cite{meissner} and up to $Q^{2}$$\simeq$0.35 GeV$^{2}$ 
for those of Ji \emph{et al.} done in the heavy baryon 
approximation~\cite{Ji chipt}. The range of validity of the $\chi PT$ 
calculations seems 
larger than of individual nucleons~\cite{review},~\cite{EG1b moments}
possibly because the $\Delta_{1232}$ resonance is suppressed in the 
Bjorken integral~\cite{Burkert Delta}. This result, however, is not 
trivial: Good agreement was expected between $\delta_{LT}$ and $\chi PT$
results since the $\Delta_{1232}$ is strongly suppressed at low $Q^2$ 
for $\delta_{LT}$. However, its measurement for the 
neutron~\cite{E94010 delta_lt} disagrees strongly with $\chi PT$ 
calculations. 

To quantitatively compare with $\chi PT$ calculations, we fit our results
up to a maximum $Q^2$ ranging from 0.30 to 0.50 GeV$^2$ (fits on lower 
$Q^2$ ranges are imprecise, and higher $Q^2$ data may lie out of the 
region  of validity for $\chi PT$). We included the data from 
Ref.~\cite{EG1a/E94010} in the fit. Our fit form is:  
\vspace{-0.2cm}
\begin{eqnarray}
\label{eq:fit}
\Gamma_{1}^{p-n}=\frac{\kappa_n^2-\kappa_p^2}{8M}Q^2+aQ^4+bQ^6
\end{eqnarray}
in which $\kappa$ is the anomalous moment of the nucleon and $a$ and 
$b$ are fit parameters. The first term in Eq.~\ref{eq:fit} stems from 
the Gerasimov-Drell-Hearn sum rule~\cite{review}. We find 
$a=0.80 \pm 0.07(stat) \pm 0.23(syst)$ and 
$b=-1.13 \pm 0.16(stat) \pm 0.39(syst)$ with $\chi ^2/dof = 1.50$. 
The $Q^4$ term agrees well with the results from Ji \emph{et al.}  
($a=0.74$) but not with those of Bernard \emph{et al.} ($a=2.4$).  The 
fit underscores the importance of the $Q^6$ term (not calculated yet in 
$\chi PT$). This was also noticed for $\Gamma_{1}^{p}$ and 
$\Gamma_{1}^{d}$~\cite{EG1b moments}.

At high $Q^{2}$, the leading twist pQCD calculation is
given by the bracket term of Eq.~\ref{eq:bj(Q2)} and is represented 
by the gray band in Fig.~\ref{fig:bjsr}. It agrees reasonably well with
the data. This implies that the total higher twist contribution
is relatively small even down to $Q^{2} \approx 1$  GeV$^{2}$ where
one would expect higher twist contributions to be significant. 
Higher twists, which measure parton
correlations, are weighted by $1/Q^{(t-2)}$ (with $t$ being the twist 
number) and are related to the confinement mechanisms and to scattering 
off coherent quarks. Because of these reasons, it was initially expected 
that higher twists would play an important role at $Q^2 \lesssim 1$ 
$GeV^2$. Higher twists can be positive or negative but there is no 
fundamental reason to expect a well-tuned cancellation of different 
terms in the higher twist series that would make the overall higher 
twist contribution small. However, this seems to be the case 
experimentally, at least around $Q^2 \approx 1$ GeV$^2$. One of the 
aims of the higher twist analysis reported here is to 
establish whether higher twists are intrinsically small, or 
whether the terms in the higher twist series conspire to cancel.

\section{Higher Twist analysis}
The first higher 
twist correction term in Eq.~\ref{eq:bj(Q2)} is~\cite{shuryak}:
\vspace{-0.2cm}
\begin{eqnarray}
\mu_{4}^{p-n} & = & \frac{M^{2}}{9}\left(a_{2}^{p-n}+4d_{2}^{p-n}+4f_{2}^{p-n}\right),
\label{eq:HT4}
\end{eqnarray}
where $a_{2}$ and $d_{2}$ 
are known. They are given by moments of the leading twist part of
$g_1$ and the twists 2 and 3 parts of $g_2$: 
$a_{2}=\int_{0}^{1}{dx\,(x}^{2}g_{1})$ and 
$d_{2} = \int_{0}^{1}dx~x^{2}\left(2g_{1}+3g_{2}\right)$. The twist-4 
term that we wish to extract is $f_{2}^{p-n}$.

\begin{figure}[ht!]
\begin{center}
\vspace*{-1.0cm}
\centerline{\includegraphics[scale=0.42, angle=0]{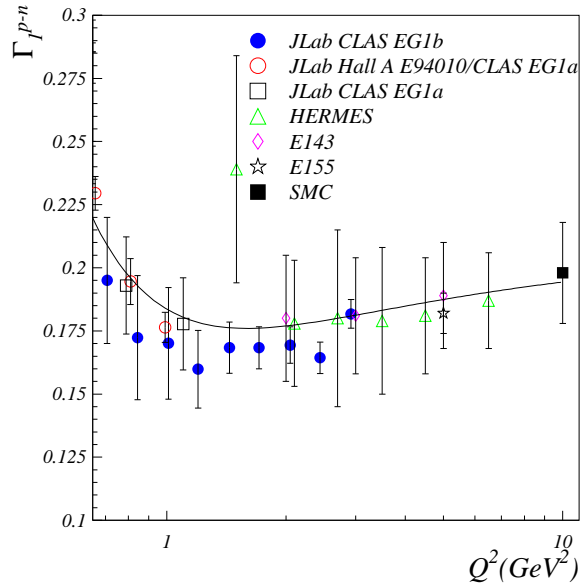}}
\end{center}
\vspace{-1.0cm}
\caption{(Color online) World data on the Bjorken integral, including 
the elastic contribution. The error bars represent the quadratic sum
of the statistic and point-to-point-uncorrelated systematic 
uncertainties for the JLab data, and the quadratic sum
of the statistic and full systematic uncertainties for the rest of the
data. The continuous line is our three parameter fit in the $Q^2$-range
from 0.66 to 10 GeV$^2$.}
\label{fig:bjHT}
\end{figure}

\begin{figure}[ht!]
\begin{center}
\centerline{\includegraphics[scale=0.46, angle=0]{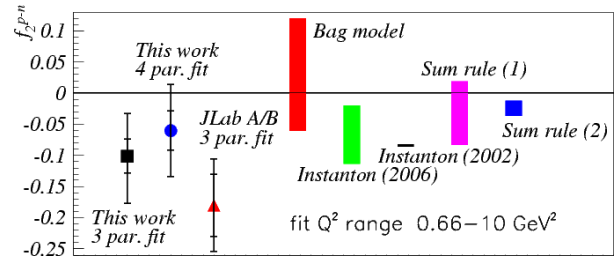}}
\end{center}
\vspace{-0.7cm}
\caption{ $f_{2}^{p-n}(Q^{2}=1$
GeV$^{2})$ for the fits performed over the $0.66<Q^{2}<10$ GeV$^2$ range 
for this study and Ref.~\cite{EG1a/E94010} (JLab A/B). 
Calculations~\cite{stein}-\cite{Weiss 2} are shown by the bands. 
Sum rule (1) refers to Ref~\cite{BBK} and (2) to Ref~\cite{stein}.}
\vspace{-0.2cm}
\label{fig:f2}
\end{figure}

To perform a higher twist analysis, the elastic contribution ($x=1$)
to $\Gamma_{1}^{p-n}$ is added. The moment $\Gamma_1^{p-n}$ which 
includes the elastic contribution estimated from form
factor parameterizations~\cite{Mergell} is shown in Fig. ~\ref{fig:bjHT}.
In Eq.~\ref{eq:bj(Q2)}, $\alpha_{s}$ is computed up to next to leading 
order.
A fit of polarized quark distributions~\cite{BB pdf} yields
$a_{2}^{p-n}=0.031\pm0.010$ at $Q^{2}$= 1 GeV$^{2}$, whereas 
$d_{2}^{p-n}=-0.007\pm0.010$ is obtained from Ref.~\cite{E155-E155x} 
and Ref.~\cite{a1n e99117} evolved to 1 GeV$^{2}$.
The EG1b data on $\Gamma_{1}^{p-n}$, together with the world's data,
can then be fit to extract $f_{2}^{p-n}$ using Eqs.~\ref{eq:bj(Q2)}
and~\ref{eq:HT4}. To account for twists
greater than rank 4, we include a coefficient $\mu_{6}^{p-n}/Q^{4}$. 
For consistency, former data on $\Gamma_{1}^{p-n}$ 
were reanalyzed using the same model as used in this paper to 
extrapolate to low $x$. For both JLab data sets 
(Ref.~\cite{EG1a/E94010} and the present data), the point-to-point
correlated uncertainties have been separated from the uncorrelated
ones. The latter are added in quadrature to the statistical 
uncertainties. The correlated systematics are propagated independently, 
as is the uncertainty arising from $\alpha_{s}$. 
The result of the fit done in the $Q^{2}$-range from 0.66 to 10.0 
GeV$^{2}$ is $f_{2}^{p-n}(Q^2=1$ GeV$^2)=-0.101\pm0.027
\pm_{0.071}^{0.063}$ with 
$\mu_{6}/M^{4}=0.084\pm0.011\pm_{0.026}^{0.022}$.
The first uncertainty is the quadratic sum of the statistical and
the point-to-point uncorrelated uncertainties. The second one
is the point to point correlated uncertainty. Comparing the values 
of $f_2^{p-n}$,  $a_{2}^{p-n}$ and $d_{2}^{p-n}$ at $Q^{2}$= 1 
GeV$^{2}$, we see that $\mu_4^{p-n} \approx 0.4 f_2^{p-n}$ GeV$^2$. 
The result for $f_2^{p-n}$ is plotted in Fig.~\ref{fig:f2} (square) 
along with the result from Ref.~\cite{EG1a/E94010} (triangle)
and theoretical predictions (In addition to $f_2$ and $\mu_{6}$, 
the third fit parameter mentioned in Figs.~\ref{fig:bjHT} 
and~\ref{fig:f2} is $g_a$, which 
was free to vary within its experimental uncertainty). As discussed 
in the introduction, only the $Q^2$-dependence of $f_2$ is known from 
pQCD. The absolute value can be computed solely from non-perturbative 
means and is difficult to obtain with Lattice QCD. For these reasons, 
only phenomenological models are available for comparison with our 
results.

At $Q^{2} = 1$ GeV$^{2}$, the leading twist term $\mu_{2}^{p-n}$ and 
higher twist terms $\mu_{4}^{p-n}$ and $\mu_{6}^{p-n}$ are of similar 
sizes but with alternating signs and with $\mu_{4}^{p-n}$ and 
$\mu_{6}^{p-n}$ mostly canceling each other.

To study the systematics associated with this higher twist analysis and
to check the legitimacy of our procedure at low $Q^2$, we conducted
several tests:
\begin{enumerate}
\item We repeated the fit for several $Q^{2}$ ranges;
\item We reiterated this work adding a $\mu_{8}^{p-n}/Q^{6}$ term to 
study the convergence of the twist series (the resulting $f_2^{p-n}$ is 
shown in Fig.~\ref{fig:f2} by the solid circle);
\item We investigated the dependence on the low $x$ extrapolation 
using different Regge-based parameterizations;
\item We extensively studied the stability of the fit for different 
choices of number of parameters and of $Q^{2}$ ranges by using 
different models that reproduce the data reasonably well. We used 
ranges from $0.47<Q^{2}<10$ to $3<Q^{2}<10$  GeV$^{2}$ and we fit with
functional forms with highest term from $\mu_{6}/Q^{4}$ 
to $\mu_{12}/Q^{10}$. 
\end{enumerate}
All observations supports the validity of our extractions. See Ref.~\cite{EG1b bjorken note} for details.

\section{Color polarizabilities}
Combination of higher twist coefficients can be  interpreted
in terms of color polarizabilities, which describe the response
of the color magnetic and electric fields to the spin of the nucleon.
The color electric and magnetic 
polarizabilities~\cite{stein},~\cite{Ji col. pol.} are 
$\chi_{E}=\frac{2}{3}\left(2d_{2}+f_{2}\right)$
and $\chi_{B}=\frac{1}{3}\left(4d_{2}-f_{2}\right)$. Using the value
of $f_{2}^{p-n}$ extracted from the fit with $Q_{min}^{2}=0.66$,
we obtain $\chi_{E}^{p-n}=-0.077\pm0.050$ and 
$\chi_{B}^{p-n}=0.024\pm0.028$.
The point-to-point correlated and uncorrelated uncertainties on $f_{2}$
were added in quadrature. Our higher twist analysis yields 
${|f}_{2}^{p-n}|\gg|d_{2}^{p-n}|$ (a feature predicted by 
models~\cite{Weiss 1} and~\cite{Weiss 2}). Consequently 
$\chi_{E}^{p-n}\simeq\frac{2}{3}f_{2}^{p-n}$ and
$\chi_{B}^{p-n}\simeq-\frac{1}{3}f_{2}^{p-n}$.

\section{Electromagnetic polarizability}
We now turn to the generalized forward spin polarizability $\gamma_{0}$. 
Spin polarizabilities characterize  the coherent response of the 
nucleon to photons. They are defined using low-energy theorems in the form 
of a series expansion in the photon energy.  The first term of the 
series comes from the spatial distribution of charge and current (form 
factors) while the second term results from the deformation of these 
distributions induced by the photon (polarizabilities). Hence, 
polarizabilities are as important as form factors in understanding 
coherent nucleon structure. \emph{Generalized} spin polarizabilities 
describe the response to \emph{virtual} photons. The low energy theorem 
defining the generalized forward spin polarizability is:
\begin{eqnarray}
\Re e[g_{TT}(\nu,Q^{2})-g_{TT}^{p\hat{o}le}(\nu,Q^{2})]= \label{eq:sr2}  \\
(\frac{2\alpha}{M^{2}})I_{TT}(Q^{2})\nu+\gamma_{o}(Q^{2})\nu^{3}+O(\nu^{5}),\nonumber
\end{eqnarray}
where $g_{TT}$ is the spin-flip doubly-virtual Compton scattering 
amplitude, and $I_{TT}$ is the coefficient of the $O(\nu)$ term of the 
Compton amplitude which can be used to generalize the 
Gerasimov-Drell-Hearn (GDH) sum rule to non-zero 
$Q^2$~\cite{GDH,review}. We have $I_{TT}(Q^{2}=0)=\kappa /4$. In 
practice $\gamma_{0}$ can be obtained from a sum rule which has a 
derivation akin to that of the GDH sum rule:
\begin{eqnarray}
\gamma_{0}=\frac{16\alpha M^{2}}{Q^{6}}\int_{0}^{x_{0}}x^{2}\left(g_{1}-
\frac{4M^{2}}{Q^{2}}x^{2}g_{2}\right)dx,
\end{eqnarray}
where $g_2$ is the second spin structure function and $\alpha$ is the fine
structure constant. Similar relations define the generalized 
longitudinal-transverse polarizability $\delta_{LT}$ :
\begin{eqnarray}
\Re e[g_{LT}(\nu,Q^{2})-g_{LT}^{p\hat{o}le}(\nu,Q^{2})]=\\
(\frac{2\alpha}{M^{2}})QI_{LT}(Q^{2})+Q\delta_{LT}(Q^{2})\nu^{2}+O(\nu^{4}),\nonumber
\end{eqnarray}
\begin{eqnarray}
\delta_{LT} = \frac{16\alpha M^{2}}{Q^{6}}\int_{0}^{x_{0}}x^2\left( g_{1}+g_{2} \right)dx.\label{eq:srdlt}
\end{eqnarray}
where $g_{LT}$ is the longitudinal-transverse interference
amplitude, and $I_{LT}$ is the coefficient of the $O(\nu)$ term of the 
Compton amplitude. Details on the derivation of 
Eqs.~\ref{eq:sr2}-\ref{eq:srdlt} can be found in~\cite{review} 
and~\cite{rev. disp. relat.}. The isovector quantity 
$\gamma_{0}^{p}-\gamma_{0}^{n}$ eliminates the $\Delta_{1232}$ 
resonance contribution~\cite{Burkert Delta}, and therefore offers the 
same advantage as $\delta_{LT}$ when comparing to calculations based on 
$\chi PT$. Higher 
moments are advantageous because they are essentially free of the 
uncertainty associated with the low $x$ extrapolation. An isospin 
separation of $\delta_{LT}$ or $\gamma_{0}$ may help us to understand 
why the $\chi PT$ calculations fail to describe them. For example, 
the $t$-channel 
exchange of axial-vector mesons (short range interactions), which are 
not included in the calculations, could be identified if one of the 
isospin components agrees with the $\chi PT$ calculations while the other disagrees.

\begin{figure}[ht!]
\begin{center}
\vspace*{-1.cm}
\centerline{\includegraphics[scale=0.45, angle=0]{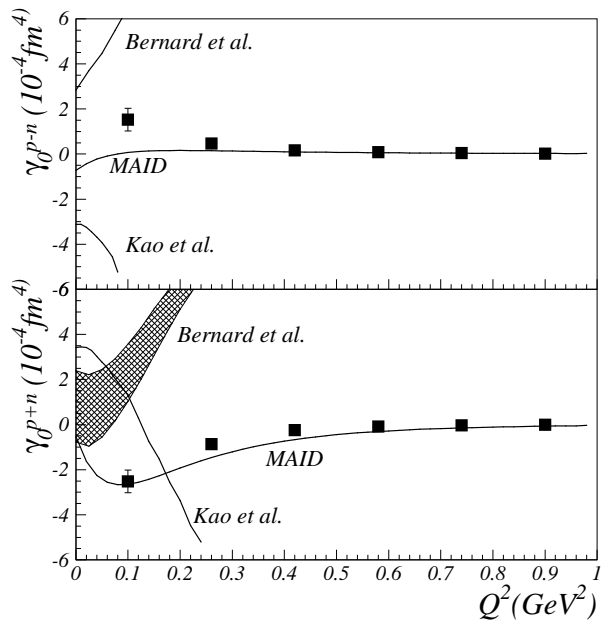}}
\end{center}
\vspace{-1.cm}
\caption{The isovector $\gamma_{0}^{p}-\gamma_{0}^{n}$ (top) 
and isoscalar $\gamma_{0}^{p}+\gamma_{0}^{n}$ (bottom) 
generalized forward spin polarizabilities  
 together with $\chi PT$-based calculations and the results from the MAID model.
The proton and neutron data are respectively from CLAS~\cite{EG1b moments} 
and Hall A~\cite{E94010 delta_lt}}
\vspace{-0.4cm}
\label{fig:gam0}
\end{figure}

We formed $\gamma_{0}^{p}-\gamma_{0}^{n}$ using the proton data from 
EG1b~\cite{EG1b moments} and the neutron data from JLab experiment 
E94010~\cite{E94010 delta_lt}. The $^{3}$He data~\cite{E94010 delta_lt} 
are more precise than the deuteron data~\cite{EG1b moments} that 
contain contributions from quasi-elastic and two-body break-up, which  
are not resolved by the CLAS spectrometer but are large at low $Q^{2}$. 
(This difficulty prevented $\gamma_{0}^{n}$ from being obtained from the
EG1b data~\cite{EG1b moments}). EG1b goes to lower $Q^{2}$ than E94010, 
but the coverage of E94010 is sufficient for our investigation.
The resulting $\gamma_{0}^{p}-\gamma_{0}^{n}$ is shown in 
Fig.~\ref{fig:gam0} (top plot) together with the predictions from 
Bernard \emph{et al.} at $O(P^{4})$~\cite{meissner} and Kao
 \emph{et al.} at $O(P^{4})$~\cite{Kao}. Experimental values are 
given in Table II. We also plot the result from the 2003 MAID 
model~\cite{MAID}. As is true for $\gamma_{0}^{p}$~\cite{EG1b moments} 
and $\gamma_{0}^{n}$~\cite{E94010 delta_lt}, $\chi PT$ calculations 
disagree with $\gamma_{0}^{p-n}$ as well. Clearly, the discrepancy seen
for $\gamma_{0}^{p}$ and $\gamma_{0}^{n}$ cannot solely be due to the
$\Delta_{1232}$ resonance. The MAID model, which provides a relatively
good description of $\gamma_{0}^{p}$ and $\gamma_{0}^{n}$, disagrees 
mildly for their difference at the lowest $Q^2$ point. Complementary 
to this study, we formed the isoscalar part 
$\gamma_{0}^{p}+\gamma_{0}^{n}$ and compared it to the data 
(Fig.~\ref{fig:gam0} bottom plot). The gray band on the Bernard 
\emph{et al.} result is due to the uncertainty from the $\Delta_{1232}$ 
resonance. The MAID model provides a good description, whereas the 
$\chi PT$-based calculations still disagree. A disagreement in the 
$\chi PT$ 
calculation of one of the isospin components of $\gamma_{0}$ along with 
agreement for the other component might have allowed us to identify a 
missing piece, such as for example a short range interaction due to
heavy mesons, in the $\chi PT$ calculations.
However, the discrepancy between data and $\chi PT$ calculations for 
both isospin components does not allow us to draw such conclusion. 
This suggests that the non-resonant background is responsible.

\begin{table}

\begin{tabular}{|c|c|c|c|c|}
\hline 
$Q^{2}$ (GeV$^{2}$)&
$\gamma_{0}^{p-n}$ &
$\gamma_{0}^{p+n}$&
Stat.&
Syst.\tabularnewline
\hline
\hline 
0.1&
1.53&
-2.51&
$0.120$&
$0.490$\tabularnewline
\hline 
0.26&
0.470&
-0.869&
$0.021$&
$0.177$\tabularnewline
\hline 
0.42&
0.159&
-0.241&
$0.006$&
$0.058$\tabularnewline
\hline 
0.58&
0.0835&
-0.0845&
$0.0040$&
$0.0233$\tabularnewline
\hline 
0.74&
0.0441&
-0.0299&
$0.0037$&
$0.0090$\tabularnewline
\hline 
0.9&
0.0217&
-0.0103&
$0.0016$&
$0.0040$\tabularnewline
\hline
\end{tabular}
\caption{Isovector and isoscalar parts of the generalized forward spin polarizability $\gamma_0$.}
\end{table}

\section{Summary and conclusion}
The Bjorken integral  was extracted from polarized proton and 
deuteron data for $0.054<Q^{2}<2.92$ GeV$^{2}$. 
The results for intermediate $Q^{2}$ (the parton to hadron transition 
domain) are consistent  with previous JLab data in which the neutron 
information was extracted from polarized $^3$He. This region exhibits a
strong $Q^{2}$-behavior, both from pQCD evolution and from some 
higher-twist effects. On the other hand, in the high-$Q^2$ 
domain the Bjorken integral is rather flat. The data together with
kinematic constraints at $Q^2 \rightarrow 0$ also suggest a small
$Q^2$-dependence, in qualitative agreement with the generalized GDH 
sum predictions.

At the lowest $Q^2$ accessed by our data, $\chi PT$ calculations agree 
better with the Bjorken integral (an isovector quantity in which the
$\Delta_{1232}$ resonance does not contribute) than with moments on 
individual nucleons. This is not trivial since the $\chi PT$ 
calculations fail to 
describe the generalized spin polarizability $\delta_{LT}$ in which 
the $\Delta_{1232}$ is also suppressed. 

Data on the generalized forward spin polarizability $\gamma_{0}^{p-n}$ 
are not reproduced by the $\chi PT$-based calculations even though the 
$\Delta_{1232}$ does not contribute.
 
It is clear from previously published data on $\delta_{LT}$ and our 
analysis of $\gamma_{0}$ that the $\Delta_{1232}$ resonance 
contribution is not responsible for the discrepancy between data and 
calculations.  
The discrepancy between the $\chi PT$ calculations and the data occurs in all isospin 
channels, which makes it less likely that it is due to the contribution
from heavier mesons in the chiral expansion. 

The low $Q^2$ $\chi PT$ regime has been recently mapped by two 
additional dedicated experiments in CLAS using polarized 
proton~\cite{EG4p} and 
deuteron targets~\cite{EG4D} and one in Hall A using 
polarized $^3$He~\cite{E97110}. These
experiments will provide further precision tests of $\chi PT$ calculation
techniques.

The moderate  $Q^{2}$ data (1 to 3 GeV$^2$) allow us to extract higher
twist contributions and color polarizabilities. The twist-4 coefficient 
was found to be large: $f_{2}^{p-n}\simeq-0.1$ at $Q^{2}=1$ GeV$^{2}$ 
(compare to $\Gamma_{1}^{p-n}=0.125$, $a_{2}^{p-n}=0.031$ and 
$d_{2}^{p-n}=-0.007$). The uncertainty on $f_2^{p-n}$ remains relatively 
large ($\approx 70\%$); however, we have completed several 
systematic studies both with the existing data as well as simulated data
(with no statistic fluctuations) that indicate our result is 
stable. The sign and magnitude of $f_{2}^{p-n}$ agree with a recent 
analysis performed on $g_{1}$ directly~\cite{LSS-HT}. The observation 
that higher twist 
effects on $\Gamma_{1}^{p-n}$ are small overall does not imply that the 
net higher twist effect on the structure function $g_1^{p-n}$ is small 
at any $x$. It is important to study the $x$-dependence of the higher 
twists, as is done in Ref.~\cite{LSS-HT}. That $|f_2|$ is significantly 
larger than $d_2$, and that $f_2 <0$, agrees well with the prediction of 
the two-scale 
model~\cite{Weiss 2}. Overall the net effect of higher twists is small, 
because of a cancellation between the twist 4 and twist 6 terms that 
are of similar sizes but opposite signs. This trend has also been seen 
for higher twist analyses done on the unpolarized structure function 
$F_2$~\cite{osipenko}. This can be interpreted within 
a vector dominance framework: the oscillating signs arise from the 
development in series of the vector meson propagator 
$\propto 1/(Q^2-M_m^2)$ where $M_m$ is the meson mass. 

This work is supported by the U.S. Department of Energy (DOE) and
the U.S. National Science Foundation. The Jefferson Science Associates
operate the Thomas Jefferson National Accelerator Facility for the
DOE under contract DE-AC05-84ER40150.

\vskip .1truein


\end{document}